\pdfoutput=1

\newcommand{\ybco}[1]{YBa$_2$Cu$_3$O$_{#1}$}
\newcommand{\tbcod}{Tl$_2$Ba$_2$CuO$_{6+\delta}$}

\documentclass[twocolumn,preprintnumbers,amsmath,amssymb,superscriptaddress,aps,prl]{revtex4-1}

\usepackage{graphicx}
\usepackage{color}
\usepackage{hyperref}

\definecolor{orange}{rgb}{1,0.647,0}
\begin{document}

\title{Logarithmic Upturn in Low-Temperature Electronic Transport as a Signature of \\
d-Wave Order in Cuprate Superconductors}


\author{Xiaoqing Zhou}
\email{xiaoqing.zhou@colorado.edu}
\affiliation{Department of Physics, Simon Fraser University, Burnaby, BC, V5A~1S6, Canada}
\affiliation{Department of Physics, University of Colorado, Boulder 80309-0390, USA}
\author{D.~C.~Peets}
\email{dpeets@fudan.edu.cn}
\affiliation{Department of Physics and Astronomy, University of British Columbia, Vancouver, BC, V6T~1Z1, Canada}
\affiliation{State Key Laboratory of Surface Physics, Department of Physics, and Advanced Materials Laboratory, Fudan University, Shanghai 200438, People's Republic of China}
\author{Benjamin Morgan}
\affiliation{Cavendish Laboratory, Madingley Road, Cambridge, CB3 0HE, United Kingdom}
\author{W.~A.~Huttema}
\author{N.~C.~Murphy}
\author{E.~Thewalt}
\author{C.~J.~S.~Truncik}
\affiliation{Department of Physics, Simon Fraser University, Burnaby, BC, V5A~1S6, Canada}
\author{P.~J.~Turner}
\author{A.~J.~Koenig}
\affiliation{Department of Physics, Simon Fraser University, Burnaby, BC, V5A~1S6, Canada}
\author{J.~R.~Waldram}
\affiliation{Cavendish Laboratory, Madingley Road, Cambridge, CB3 0HE, United Kingdom}
\author{A.~Hosseini}
\affiliation{Department of Physics and Astronomy, University of British Columbia, Vancouver, BC, V6T~1Z1, Canada}
\author{Ruixing Liang}
\author{D.~A.~Bonn}
\author{W.~N.~Hardy}
\affiliation{Department of Physics and Astronomy, University of British Columbia, Vancouver, BC, V6T~1Z1, Canada}
\affiliation{Canadian Institute for Advanced Research, Toronto, Ontario, MG5 1Z8, Canada}
\author{D.~M.~Broun}
\email{dbroun@sfu.ca}
\affiliation{Department of Physics, Simon Fraser University, Burnaby, BC, V5A~1S6, Canada}
\affiliation{Canadian Institute for Advanced Research, Toronto, Ontario, MG5 1Z8, Canada}


\begin{abstract}
In cuprate superconductors, high magnetic fields have been used extensively to suppress superconductivity and expose the underlying normal state.  Early measurements revealed insulating-like behavior in underdoped material versus temperature $T$, in which resistivity increases on cooling with a puzzling $\log(1/T)$ form. We instead use microwave measurements of flux-flow resistivity in \ybco{6+y}\ and \tbcod\ to study charge transport deep inside the superconducting phase, in the low temperature and {\slshape low} field regime.  Here, the transition from metallic low-temperature resistivity ($d\rho/dT>0$) to a $\log(1/T)$ upturn persists throughout the superconducting doping range, including a regime at high carrier dopings in which the field-revealed normal-state resistivity is Fermi-liquid-like. The $\log(1/T)$ form is thus likely a signature of $d$-wave superconducting order, and the field-revealed normal state's $\log(1/T)$ resistivity may indicate the free-flux-flow regime of a phase-disordered $d$-wave superconductor.

\end{abstract}
                                                            
\maketitle

One of the most important conceptual issues in the copper-oxide superconductors concerns whether the pseudogap region in the underdoped part of the doping--temperature phase diagram \cite{Timusk:1999p422} contains a quantum-disordered superconductor \cite{Ruckenstein:1987gr,Baskaran:1987fk,Emery:1995p364,Franz:1998p949,FRANZ:2001co,Franz:2002p704,Herbut:2002p612,herbut02a}, with local superconducting pairing but no long-range phase coherence. Experiments in high magnetic field have played a central role in trying to settle this issue.  Early measurements on La$_{2-x}$Sr$_x$CuO$_4$ using 60~T pulsed fields to suppress superconductivity revealed 
unusual low temperature behavior characterized by a $\log(1/T)$ resistivity \cite{Ando:1995p148}, where $T$ is absolute temperature, which has been widely interpreted as an insulating response \cite{Imada:1998er}. Subsequent measurements showed that the $\log(1/T)$ behavior was confined to the underdoped side ($x < 0.16$) of the phase diagram, with metallic behavior observed on the overdoped side ($x > 0.16$) \cite{Boebinger:1996p147}. This has been taken as evidence for the existence of a quantum critical point \cite{Tallon:2001p456}, a contentious issue in the cuprate phase diagram, but a central organizing principle in many other correlated electron systems \cite{Sachdev:2000p576,Kopp:2005p742,Coleman:2005p454,Gegenwart:2008p465,Broun:2008ci,Rowley:2014bd,Kuo:2016iw}.  

A heavily phase-disordered superconductor is expected to enter a resistive regime in which a liquid of rapidly fluctuating vortex--anti-vortex pairs destroys the superconducting phase coherence, eventually forming an electronic crystal \cite{Emery:1995p364,Franz:1998p949,FRANZ:2001co,Franz:2002p704,Herbut:2002p612,herbut02a}. A key question for the cuprates is the extent to which signatures of these vortex fluctuations have been observed, with key data coming from the Nernst effect and fluctuation diamagnetism \cite{ Corson:1999p716, Xu:2000p609,Wang:2001ix,Wang:2002p645,Wang:2003cm,Ong:2004hy,Wang:2005p2400,Wang:2006p185,Li:2007p276,Sonier:2008p613,Li:2010ic}.  However, the interpretation of these measurements remains highly controversial, with an alternative body of work focussing on the quasiparticle mechanisms for Nernst effect that occur in systems with small Fermi energy \cite{Livanov:1999jy,Behnia:2009cw,Chang:2010ic,DoironLeyraud:2013kga,Behnia:2016fc,CyrChoiniere:2018ed}. The recent observation of pair density waves by scanning tunneling microscopy (STM) \cite{Hamidian:vm,Edkins:2018wo} takes this debate in a new direction, potentially providing common ground for the previously opposing points of view: as well as indicating the persistence of short-range Cooper pairing outside the superconducting phase, the observed pair density wave naturally gives rise to Fermi-surface reconstructions of the kind present in the underdoped cuprates \cite{Zelli:2012ka,Norman:2018gba}. The electrical transport signature of the pair density wave or the vortex liquid, on the other hand, has yet to be identified.

Direct current (DC) electrical transport in a heavily phase-disordered superconductor, with or without external fields, is very hard to distinguish from that of a normal metal, because moving vortices exhibit a flux-flow resistivity $\rho_\mathrm{ff}$  \cite{Bardeen:1965p151,Mackenzie:1996p199,Geshkenbein:1998p3010}.   In conventional superconductors, $\rho_\mathrm{ff}$ is closely connected to the normal state resistivity $\rho_n$ via the Bardeen-Stephen relation,  $\rho_\mathrm{ff} \sim \rho_\mathrm{n} B/B_{\mathrm{c}2}$ \cite{Bardeen:1965p151}, where $B_{\mathrm{c}2}$ is the upper critical field.
However, it remains unclear whether the Bardeen--Stephen picture  applies to the cuprates \cite{Zhou:2013if}, in which the vortex cores contain at most a few discrete states \cite{MaggioAprile:1995p3014} and $d$-wave pairing \cite{HARDY:1993p632,Shen:1993vt,Wollman:1993bs,Tsuei:1994ci} produces a substantial density of extended quasiparticle states outside the cores (e.g., populated by the Doppler-shift effect \cite{VOLOVIK:1993p201}, even at zero temperature).

\begin{figure*}[t]
\centering
\includegraphics[width= \textwidth]{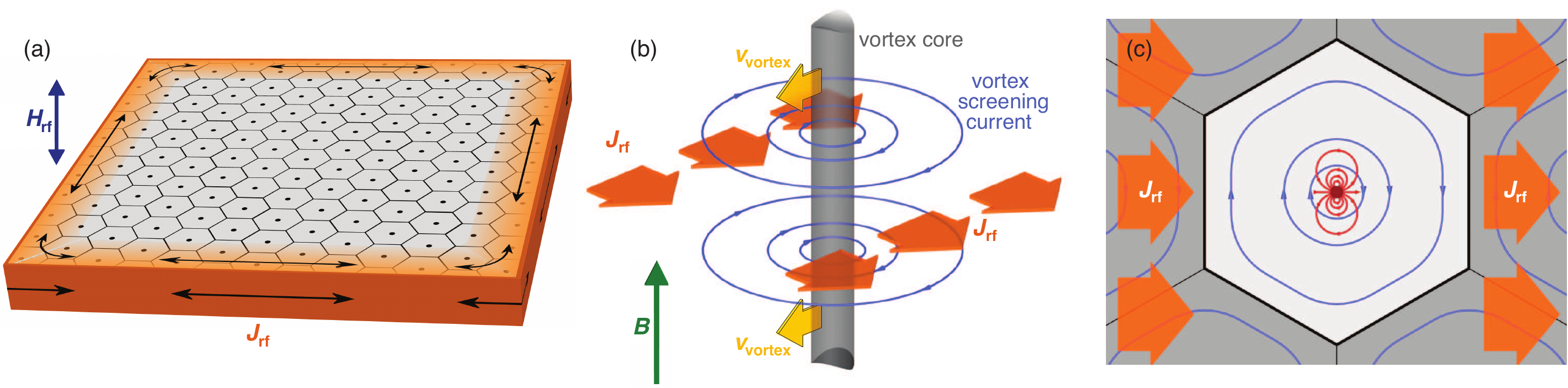}
\caption{Measuring flux-flow resistivity in cuprate superconductors.  (a) A platelet crystal is cooled in a static magnetic field $B$, perpendicular to the CuO$_2$ planes, setting up a uniform vortex lattice with in-plane screening currents.  A weak microwave magnetic field $H_\mathrm{rf}\parallel B$ induces an electrical current density $J_\mathrm{rf}$, concentrated near the edges of the platelet. (b) A superconducting flux line experiences a Lorentz force $\Phi_0 J_\mathrm{rf}$ per unit length, where $\Phi_0$ is the superconducting flux quantum, driving the vortex sideways with velocity $v_\mathrm{vortex}$ (yellow arrows).  By stimulating the vortex at microwave frequencies and measuring the vortex velocity in a phase-sensitive manner, both dissipative and elastic forces on the flux line can be resolved. (c) Over most of the vortex lattice unit cell, power absorption is little changed from its value in the absence of vortices.  However, near the center of the unit cell, the oscillatory motion of the vortex induces additional electric fields (red), which couple to charge excitations in the vortex core giving rise to increased dissipation, parameterized by the vortex resistivity $\tilde \rho_\mathrm{v}$, as described in Supplemental Material S1B.}
\label{fig1} 
\end{figure*}

To provide new insights into these issues, in this Letter we report electrical transport measurements deep inside the superconducting state, in the low-field, low temperature limit, where the existence of vortices is unambiguous and the inter-vortex spacing is much greater than the vortex core size.  We use high quality single crystals of \ybco{6+y} and \tbcod\  (see below and Supplemental Material) to cross the doping--temperature phase diagram using a trajectory in field and doping inaccessible to high-field DC resistivity measurements. In this regime the vortices are pinned by defects, making their dissipative response extremely difficult to probe in DC experiments since exceedingly large current densities are required to initiate free flux flow. We instead perform a perturbative, phase-sensitive measurement at microwave frequency to infer both the vortex pinning (see Supplemental Material) and flux-flow resistivity from the complex electrical response of the flux lines \cite{JIGittleman:1968p172,COFFEY:1991p156,BRANDT:1991p149}. The flux-flow resistivity at all dopings follows a universal, $\log(1/T)$ temperature dependence, even on the overdoped side where the field-revealed normal-state resistivity is Fermi-liquid-like. Since this rules out a Bardeen--Stephen-type mechanism, in which resistive losses in the cores simply mimic the underlying normal state, the $\log(1/T)$ behavior is evidently an intrinsic property of cuprate vortices, suggesting that the field-revealed $\log(1/T)$ resistivity in the underdoped regime may indicate free flux flow.

\begin{figure*}[t]
\centering
\includegraphics[width= \textwidth]{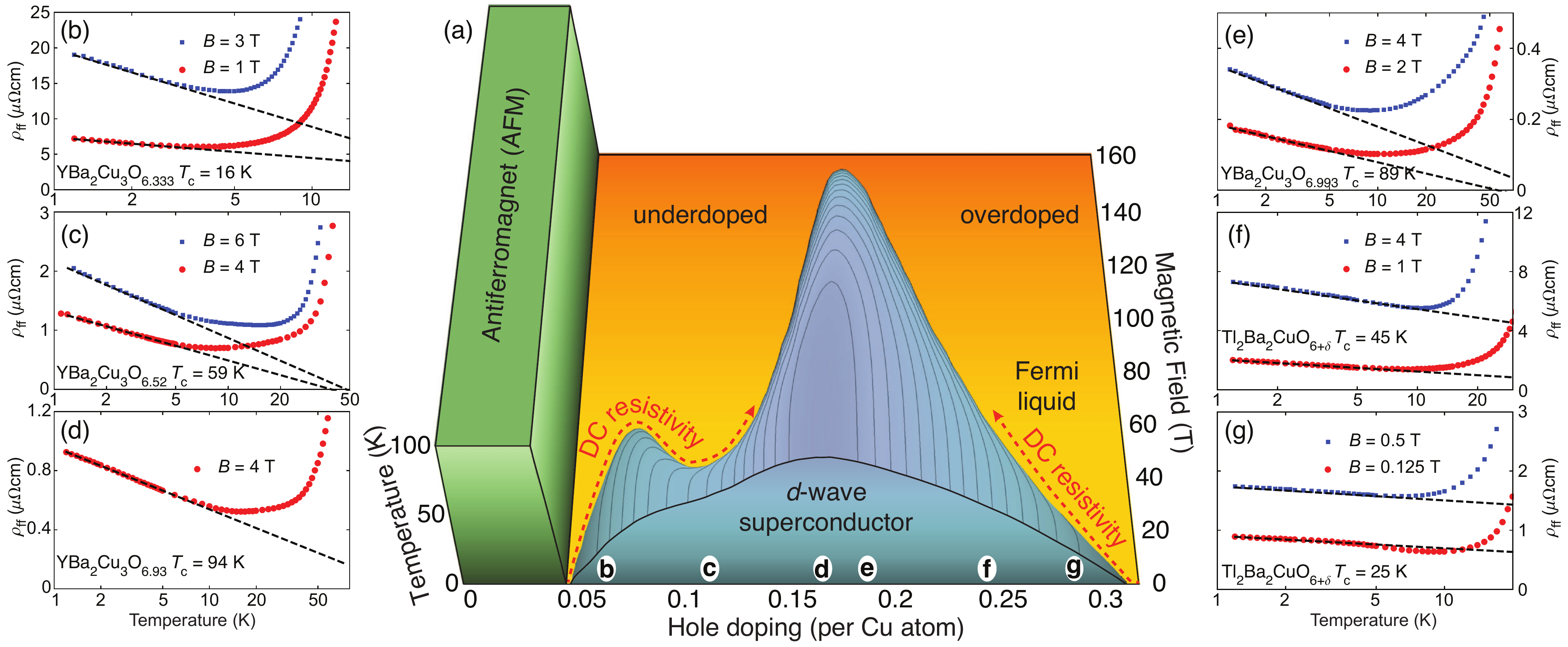}
\caption{Flux-flow resistivity across the hole-doped cuprate superconducting dome.  (a) Phase diagram, based on data from \ybco{6+y}\ \cite{Liang:2006hr} and \tbcod\ \cite{Bangura:2010p1675}.  To the left of optimal $T_\mathrm{c}$, on the underdoped side of the phase diagram, superconductivity emerges from the pseudogap regime \cite{Timusk:1999p422}. To the right of the superconducting dome, on the overdoped side of the phase diagram, the normal state is consistent with a Fermi liquid \cite{Mackenzie:1996p199,Vignolle:2008p1694,Bangura:2010p1675}, and superconductivity is BCS-like, with an energy gap that scales as $T_\mathrm{c}$ and closes in the normal state \cite{Hawthorn:2007hq,Huefner:2008p315,LeeHone:2017to,LeeHone:2018ug}.  Surrounding the phase diagram are semi-log plots of low-temperature flux-flow resistivity data:  (b) \ybco{6.333}, $T_\mathrm{c} = 16$~K; (c) Ortho-II ordered \ybco{6.52}, $T_\mathrm{c} = 59$~K \cite{Zhou:2013if}; (d) optimally doped \ybco{6.93}, $T_\mathrm{c} = 94$~K; (e) Ortho-I ordered \ybco{6.993}, $T_\mathrm{c} = 89$~K;  (f) \tbcod, $T_\mathrm{c} = 45$~K; and (g) \tbcod, $T_\mathrm{c} = 25$~K.  Dashed lines are low temperature fits to $\rho_\mathrm{ff}(T) = \rho_0 + \rho_1 \log(1/T)$.}
\label{fig2}
\end{figure*}

The experimental concept is summarized in Fig.~\ref{fig1}. A perturbative microwave field $H_\mathrm{rf}$ induces a screening current density $J_\mathrm{rf}$ at the outer edges of the sample, with $J_\mathrm{rf}$ being nearly uniform within a vortex-lattice unit cell.  We measure the ratio of the tangential components of the electric and magnetic rf fields at the sample surface via the cavity perturbation technique \cite{Klein:1993p1129,Bonn:2007hl} (Supplemental Material S1A). Since electric field $E = \tilde \rho J$, this directly probes the complex resistivity $\tilde \rho$. In zero applied magnetic field, the Meissner response $ \tilde \rho_\mathrm{s}$ is mostly imaginary and dominated by the reactive superfluid dynamics of the superconductor, with a small density of quasiparticle excitations present that couple to the rf fields and result in a nonzero dissipative term. In field, a flux-line lattice is established and the induced vortex motion contributes a new term $\tilde \rho_\mathrm{v}$ to the electrodynamics that is approximately additive in the complex resistivity: $\tilde \rho_\mathrm{eff} \simeq \tilde \rho_\mathrm{s} + \tilde \rho_\mathrm{v}$  \cite{COFFEY:1991p156,BRANDT:1991p149}. The dynamical process is illustrated in Fig.~\ref{fig1}(b): the magnetic vortex is subject to the Lorentz force applied by the oscillating current $J_\mathrm{rf}$, and hence vibrates perpendicular to $J_\mathrm{rf}$ with an amplitude typically much less than 1~\AA\ \cite{Zhou:2013if}. The motion of the  flux line induces an electrical dipole field 90\,$^\circ$ out of phase with the reactive pinning force, which couples to quasiparticle excitations in the vicinity of the core and manifests dissipation [Fig.~\ref{fig1}(c)]. In short, the flux-line behaves as a driven damped harmonic oscillator, with the viscous friction between the vortex and the electrical fluid (i.e., flux-flow resistivity) as the damping term. As outlined in Supplemental Material S1B, the flux-flow resistivity and pinning constant can be simultaneously extracted from our phase-sensitive measurements of the complex $\tilde \rho_\mathrm{v}$. The robustness of the extraction has been empirically validated by (1) comparing our extracted flux-flow resistivity to DC  resistivity measurements; (2) comparing our extracted pinning constant to that inferred from DC magnetization measurements; and (3) comparing our results with measurements at higher microwave and terahertz frequencies (Supplemental Material S2).  

Our four \ybco{6+y} and two \tbcod\ samples span the entire range of superconducting dopings in the cuprate phase diagram; the extracted flux-flow resistivity is plotted in Fig.~\ref{fig2}.  In each sample, $\rho_\mathrm{ff}(T)$ drops rapidly as the vortex cores shrink on cooling below $T_\mathrm{c}$.  After a minimum at an intermediate temperature, $\rho_\mathrm{ff}(T)$ rises again and, in all samples, transitions into a regime in which $\rho_\mathrm{ff}(T) \approx \rho_0 + \rho_1 \log(1/T)$.  For the most underdoped sample, \ybco{6.333}, this behavior is not surprising: its zero-field normal state resistivity is known to follow a $\log(1/T)$ form \cite{Sun:2005p195} and the Bardeen--Stephen picture would then lead us to expect the same in $\rho_\mathrm{ff}(T)$.  The observation of $\log(1/T)$ flux-flow resistivity in the more highly doped \ybco{6+y} samples is more surprising, since these materials are exceptionally clean and have a high degree of CuO chain order.  As a measure of their quality, zero-field microwave spectroscopy of similar samples of Ortho-II \ybco{6.52} and Ortho-I \ybco{6.993} indicates low temperature transport mean free paths exceeding 5~$\mu$m \cite{Turner:2003p331,Harris:2006p388}, while quantum oscillation measurements on Ortho-II \ybco{6.52} give a dephasing mean free path $\ell$ $\sim$ 350~\AA\ in the normal state, corresponding to $k_\mathrm{F} \ell > 40$, where $k_\mathrm{F}$ is the Fermi wavenumber \cite{Ramshaw:2011p2289}.  This is a regime in which strong localization should not occur.

At $T_\mathrm{c} = 89$~K (slightly overdoped) we exhaust the doping range that can be explored in pure \ybco{6+y} and switch to the \tbcod\ system.  The overdoped side differs from the underdoped region in that superconductivity is BCS-like, with an energy gap proportional to $T_\mathrm{c}$ that closes on warming through $T_\mathrm{c}$ \cite{Hawthorn:2007hq,Huefner:2008p315,LeeHone:2017to,LeeHone:2018ug}. The normal state is generally regarded as consistent with Fermi liquid theory, since quantum oscillation experiments show a large, unreconstructed Fermi surface, with Fermi-liquid temperature damping \cite{Vignolle:2008p1694,Bangura:2010p1675}. Furthermore, the field-revealed, low-temperature resistivity in as-grown material is metallic,  with power-law exponents close to the Fermi liquid value of two \cite{Mackenzie:1996p199} and a $k_\mathrm{F} \ell$ value of approximately 300 \cite{Vignolle:2008p1694,Bangura:2010p1675}. In particular, superconductivity in highly overdoped samples can be suppressed by laboratory-scale magnetic fields, yet no indication of $\log(1/T)$ upturns has been observed at low temperature \cite{Mackenzie:1996p199} (also see Supplemental Material S2D for in-field normal-state measurements on our own sample). It is therefore extremely surprising that flux-flow resistivity exhibits a $\log(1/T)$ temperature dependence in this doping range, as seen in Figs.~\ref{fig2}(f) and \ref{fig2}(g) for $T_\mathrm{c} = 45$~K and $T_\mathrm{c} = 25$~K  \tbcod, respectively.  Were the Bardeen--Stephen relation to apply, the flux-flow resistivity would decrease monotonically with decreasing temperature. The breakdown of the Bardeen--Stephen relation in overdoped material is our most significant observation, since it demonstrates that the $\log(1/T)$ flux-flow resistivity does not have its roots in the normal state, but is a property of the vortices themselves.

\begin{figure}[t]
\centering
\includegraphics[width=0.9 \columnwidth]{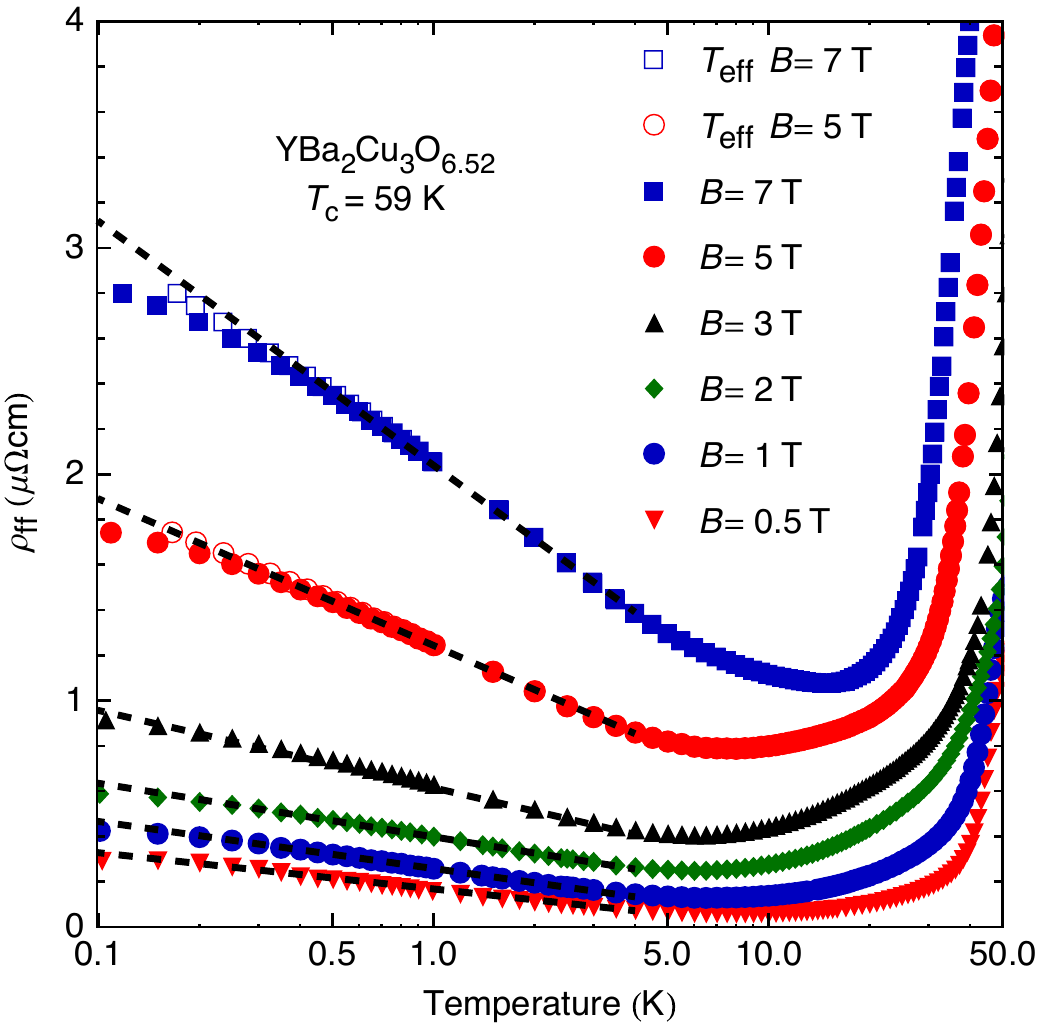}
\caption{Low temperature flux-flow resistivity versus temperature. To track the flux-flow resistivity to lower temperature, the Ortho-II \ybco{6.52} sample was remeasured in a $^3$He--$^4$He-dilution-refrigerator version of our experiment.  The $\log(1/T)$ upturn in $\rho_\mathrm{ff}(T)$ softens below 0.4~K, likely due to finite-frequency effects, as the 2.50~GHz microwave photon energy corresponds to a temperature scale $T_\omega \equiv \hbar \omega/k_\mathrm{B} = 125$~mK.  To illustrate how this may affect the measurement, at the two highest fields $\rho_\mathrm{ff}$ has also been plotted against an effective temperature $T_\mathrm{eff} \equiv \sqrt{T^2 + T_\omega^2}$ for $B = 5$~T and 7~T (open symbols).}
\label{fig3} 
\end{figure}

In-plane resistivity measurements on nonsuperconducting \ybco{6.35} found $\log(1/T)$ behavior extending down to a temperature of 80~mK \cite{Sun:2005p195}. To test whether the microwave flux-flow resistivity exhibits similar behavior, we performed a second version of our experiment in a $^3$He--$^4$He dilution refrigerator \cite{Truncik:2013hr} on the Ortho-II \ybco{6.52} sample. Low temperature flux-flow resistivity is shown in Fig.~\ref{fig3}, for fields ranging from 0.5 to 7~T.  The $\log(1/T)$ term observed at higher temperatures extends below 1~K, but begins to soften at around 0.4~K, likely a consequence of the finite measurement frequency. The energy of 2.50~GHz microwave photons corresponds to a temperature scale $T_\omega \equiv \hbar \omega/k_\mathrm{B} = 125$~mK. As a heuristic illustration of how this might impact the data, for the two highest fields we plot $\rho_\mathrm{ff}$ against an effective temperature $T_\mathrm{eff} \equiv \sqrt{T^2 + T_\omega^2}$, in Fig.~\ref{fig3} --- the $\log(1/T)$ form is recovered.  Previous measurements above 1.5\,K on \ybco{6.52} (Fig.~15 of Ref.~\onlinecite{Zhou:2013if}) support this interpretation, showing low temperature softening of $\rho_\mathrm{ff}(T)$ that becomes more pronounced with increasing frequency.

To this point, the global picture of cuprate charge transport has been based on DC resistivity measurements carried out in the field-revealed normal state \cite{Ando:1995p148,Boebinger:1996p147}. These data show a transition, near optimal doping, from $\log(1/T)$ resistivity ($d\rho/d T < 0$) in underdoped material to metallic behavior ($d\rho/dT > 0$) on the overdoped side, in a manner suggestive of a quantum critical point \cite{Ando:1995p148,Boebinger:1996p147}. Being a superconducting probe, flux-flow resistivity measurements have allowed us to traverse from the underdoped to the overdoped regime in the low-field limit, without having to detour through the field-revealed normal state, as illustrated in Fig.~\ref{fig2}(a).  Our results add to the global transport picture in several significant ways. First, we have observed that the $\log(1/T)$ flux-flow resistivity persists across the entire superconducting doping range, including in heavily overdoped \tbcod, implying that the $\log(1/T)$ behavior is intimately connected to the superconductivity. Second, we observe $\log(1/T)$ flux-flow resistivity in materials in which $k_\mathrm{F} \ell \gg 1$, ruling out disorder-induced localization as the underlying cause. We now turn to the interpretation and possible origin of this behavior.

In the Bardeen--Stephen model \cite{Bardeen:1965p151}, flux-flow resistivity is given by the normal state resistivity scaled by the vortex-core fraction: $\rho_\mathrm{ff} \sim \rho_\mathrm{n} B/B_{\mathrm{c}2}$. This form has proven very successful for conventional superconductors, as has also been confirmed in our own studies on conventional superconductors Nb, NbSe$_2$ and V$_3$Si (Supplemental Material S2A). There are two reasons why Bardeen--Stephen theory is particularly applicable to conventional superconductors: first, the vortex cores are large, and contain a near-continuum of single-particle states, approximating the normal-state density of states; and second, since the pairing symmetry is nodeless, at low temperatures the vortex cores can be treated as normal-state metallic tubes embedded in a single-particle vacuum. However, we stress that cuprates are very different \cite{BREWER:we} in these respects, since the vortex cores are small \cite{MaggioAprile:1995p3014} and the few discrete states they contain cannot be approximated by a metallic continuum. More importantly, due to the $d$-wave pairing symmetry, the vortex cores are surrounded by a gas of low-energy nodal quasiparticles.

Microwave spectroscopy of Ortho-II \ybco{6.52} has established a close connection between nodal-quasiparticle dynamics and vortex dissipation in the cuprates \cite{Zhou:2013if} (Supplemental Fig.~S7). These experiments reveal a vortex viscosity, $\eta(\omega,T) = B \Phi_0/\rho_\mathrm{ff} $, which has a frequency and temperature dependence very similar to the zero-field quasiparticle conductivity, $\sigma_1(\omega,T)$ \cite{Turner:2003p331,Harris:2006p388}. This strongly suggests that external nodal quasiparticles, rather than internal core states, are the predominant damping mechanism for vortex motion. This contrasts sharply with conventional superconductors in which the single-particle excitations are contained inside the vortex core and have a close correspondence with the states of the normal metal. In Ortho-II \ybco{6.52} \cite{Zhou:2013if}, the viscosity (and quasiparticle conductivity) exhibit pronounced peaks as a function of temperature, due to the competition between a rapidly increasing quasiparticle lifetime on cooling, and the condensation of quasiparticles into the superconducting condensate. A peak in $\eta(T)$ then leads naturally to a minimum in $\rho_\mathrm{ff}(T)$ at intermediate temperatures, followed by a low temperature upturn.  This picture explains the qualitative form of $\rho_\mathrm{ff}(T)$, but on its own does not account for the detailed $\log(1/T)$ temperature dependence of flux-flow resistivity, which remains an open question.  The connection to quasiparticle physics is nevertheless strongly suggestive and warrants further investigation.

A similar $\log(1/T)$ power law has been observed in underdoped SmFeAsO$_{1-x}$F$_x$, where it is strengthened by field\cite{Riggs2009}.  
This system is not thought to be nodal, and has magnetic rare earth ions which can interact with the applied field.  SmFeAsO$_{1-x}$F$_x$ thus offers a test for our interpretation --- since its $\log(1/T)$ term is evidently field-induced rather than field-revealed and only exists at low dopings, the $\log(1/T)$ contribution to the low-field flux-flow resistivity should be much weaker or absent, and not exist at higher dopings.  If the $\log(1/T)$ terms in flux-flow and high-field resistivity have independent origins or do not arise from $d$-wave vortices, the doping dependence of $\rho_\mathrm{ff}(T)$ in SmFeAsO$_{1-x}$F$_x$ may more closely resemble that of the cuprates.

In summary, our results indicate that $\log(1/T)$ flux-flow resistivity is an intrinsic dynamical property of cuprate vortices, which has several implications for charge transport across the phase diagram.  It explains the surprising observation of $\log(1/T)$ behavior in very clean materials for which $k_\mathrm{F} \ell \gg 1$. The striking similarity between the flux-flow resistivity, field-revealed normal-state resistivity, and the resistivity of nonsuperconducting underdoped samples also suggests that the low temperature pseudogap regime contains a vortex liquid in free flux flow, with the observation of $\log(1/T)$ resistivity outside the superconducting dome (at low temperatures, and either high fields or low hole-dopings) acting as a signature of short-range superconducting order and vortex fluctuations \cite{Corson:1999p716,Xu:2000p609,Wang:2005p2400}. The existence of such a vortex-liquid regime provides a natural explanation of Wiedemann--Franz-law violation in underdoped YBa$_2$Cu$_3$O$_{6 + y}$ \cite{DoironLeyraud:2006p2408}:  in a vortex liquid the electrical transport probes the flux-flow resistivity, with the bulk medium and the vortices contributing additively to the total dissipation \cite{COFFEY:1991p156,BRANDT:1991p149}. In contrast, thermal currents in the vortex state are carried by bulk quasiparticles, with negligible contribution from the vortex cores \cite{Kubert:1998ic}.  

We thank I.~Affleck, J.~S.~Dodge, M.~P.~Kennett and J.~E.~Sonier for useful discussions. Research support was provided by the Natural Science and Engineering Research Council of Canada (NSERC), the Canadian Foundation for Innovation (CFI), the Canadian Institute for Advanced Research (CIFAR) and the National Natural Science Foundation of China (NSFC, grant No.~11650110428).

\end{document}